\documentclass[prl,twocolumn,amssymb,amsfonts,superscriptaddress,floatfix,showpacs]{revtex4}

\usepackage[dvips]{graphicx,color}

\newcommand{\cD}{{\cal D}}

\newcommand{\cL}{{\cal L}}

\newcommand{\trace}{\mbox{tr}}

\newcommand{\mket}[1]{\vert{#1}\rangle}
\newcommand{\mbra}[1]{\langle{#1}\vert}

\begin{document}

\title{ Experimental Implementation of a Concatenated Quantum Error-Correcting Code } 

\author{Nicolas Boulant}
\affiliation{Department of Nuclear Engineering, Massachusetts Institute
of Technology, Cambridge, MA 02139, USA}
\author{Lorenza Viola}
\thanks{Corresponding author: {\tt lorenza.viola@dartmouth.edu}}
\affiliation{Department of Physics and Astronomy, Dartmouth College,
Hanover, NH 03755, USA}
\author{Evan M. Fortunato }
\affiliation{{\sc alphatech}, Inc. 6 New England Executive Park, 
Burlington, MA 01803, USA} 
\author{David G. Cory}
\affiliation{Department of Nuclear Engineering, Massachusetts Institute
of Technology, Cambridge, MA 02139, USA}

\date{\today}

\begin{abstract}
Concatenated coding provides a general strategy to achieve the desired
level of noise protection in quantum information storage and
transmission.  We report the implementation of a concatenated quantum
error-correcting code able to correct against phase errors with a
strong correlated component. The experiment was performed using
liquid-state nuclear magnetic resonance techniques on a four spin
subsystem of labeled crotonic acid. Our results show that
concatenation between active and passive quantum error correcting
codes offers a practical tool to handle realistic noise contributed by
both independent and correlated errors.
\end{abstract}

\maketitle

Scalable quantum information processing (QIP) requires the ability to
realize information fault-tolerantly in the presence of both
environmental and control errors. A variety of approaches has been
developed to meet this challenge, including active 
quantum error-correcting codes (QECCs)~\cite{qec0}, dynamical
decoupling techniques ~\cite{viola}, passive 
QECCs based on decoherence-free subspaces (DFSs) and noiseless
subsystems (NSs)~\cite{dfss,klv}, and topological
schemes~\cite{kitaev}.  While the exploration of viable routes to
quantum fault-tolerance is witnessing continual advances
(see~\cite{knill04} and~\cite{kaveh} for recent threshold analyses of
post-selected QIP and concatenated decoupling schemes, respectively),
QECCs remain to date the method of choice under a relatively wide range
of error and control assumptions~\cite{steane04}.  In particular,
concatenated QECCs are instrumental to ensure that a final accuracy
can be reached without requiring arbitrarily low error rates at
intermediate stages~\cite{knill:1996}. The basic idea is to use
multiple levels of encoding to recursively obtain logical qubits with
improved robustness. In its standard setting, a concatenated code
consists of hierarchically implementing a fixed QECC, provided that
the errors for the encoded information satisfy at each level
appropriate assumptions~\cite{knill:1996,steane04}. For the procedure
to be successful, it is critical that the implementation begins with
sufficiently high fidelity, which requires the entry-level physical
qubits to be subjected to a sufficiently weak noise.

If small error rates are not available from the start, concatenated
QECCs are still valuable if the originating noise process is highly
correlated, which makes it possible to exploit the existence of
efficient DFS or NS encodings~\cite{dfss,klv}. Because the latter are
tied to the occurrence of symmetries in the error process, such {\it
infinite-distance} QECCs are capable of tolerating arbitrarily high
error rates as long as the underlying symmetry is exact. While
infinite-distance behavior is not retained for imperfect symmetry,
stability results ensure that the residual errors remain small if the
symmetry is broken perturbatively, with short-time fidelity solely
determined by the perturbing noise strength~\cite{bacon,zanardi3}.
Concatenation schemes taking advantage of both finite- and-infinite
distance codes were originally developed in \cite{lidar2} and
subsequently \cite{hwang} in the context of the so-called cluster
error model, where a dominant collective symmetry is perturbed by
independent errors on individual qubits.

Here, we theoretically expand and experimentally demonstrate the
usefulness of concatenating active and passive QECCs.  Our approach is
tailored to realistic {\it hybrid} noise models where errors do not
follow a cluster pattern. In particular, while being to a large extent
independent, they are dominated by a large error rate which prevents
quantum error correction (QEC) from being affordable with feasible
control resources. Two guiding principles emerge for error control
design: (1) treat errors in order of their importance; (2) at each
stage, realize logical qubits with reduced error rates.  Unlike
standard concatenation schemes, where physical qubits are uniformly
replaced by logical ones at the first level of encoding, this leads in
general to effect such a replacement only partially at a given stage,
with physical and logical qubits being treated alike as needed.
 
{\it Concatenated active and passive QECCs.-} Let $S$ be the quantum
system of interest, and imagine that noise on $S$ is to a good
approximation Markovian.  Then the state of $S$ evolves as $\rho_t =
e^{{\cL}t} [\rho_0]$, where the infinitesimal noise super-operator
${\cL}$ takes the standard Lindblad form \cite{alicki},
\begin{equation}
\cL[\rho]=\sum_\mu {\cD}_{L_\mu}[\rho]= \frac{1}{2} \sum_\mu 
\Big( 
[L_\mu \rho, L_\mu^\dagger] +
[L_\mu, \rho L_\mu^\dagger] \Big) \:. 
\label{sme}
\end{equation}
Given the set of error generators $\{ L_\mu\}$, a measure of the
overall noise strength is given by $\lambda=\sum_\mu |L_\mu|^2 +
|\sum_\mu L_\mu^\dagger L_\mu|$, where
$|X|=\text{Max}|\text{eig}(\sqrt{X^\dagger X})|$~\cite{klv}. For
independent noise on qubits, each $L_\mu$ involves a single-qubit
Pauli operator, and $\lambda$ can be thought of as resulting from the
sum of the partial noise strengths $\lambda_\mu=2 |L_\mu|^2$
associated to each error generator. In general, the error probability
for information stored in $S$ is a complicated function of time. By
letting $F_e(t)$ denote entanglement fidelity \cite{schumacher}, one
may write an error expansion
\begin{equation}
F_e(t) = 1 + \sum_{k=1}^\infty {1 \over {k!}} 
\Big( {t \over \tau_k} \Big)^k \:, 
\label{taylor}
\end{equation}
where the $k$th order error rate $1/|\tau_k^k|$
is upper-bounded by $\lambda^k$. If information is protected in a
$w$-error-correcting code, error rates up to $w$th order are
effectively canceled, improving the fidelity to $F_e(t)= 1 -
O[(t/\tau_{w+1})^{w+1}] \geq 1 -O[(\lambda t)^{w+1}]$~\cite{klv}.  
The larger the window where $F_e(t)$ remains approximately flat, the
longer the time interval after which a one-time use of the code still
succeeds at retaining the information with high fidelity. Since such
an interval is determined by $\lambda$, the condition that noise
is sufficiently weak is critical for QEC.


We illustrate our error control methodology by focusing on the
following example. Consider three physical qubits with independent
phase errors. Strong noise on one of the qubits, say qubit 3, causes
the overall strength to be too high for QEC to produce a significant
improvement. However, noise on qubit 3 is dominated by correlated
dephasing involving an additional qubit, say qubit 4. The error model
may be specified in terms of the following generators:
$L_1=\sqrt{\lambda_1/2} \sigma^1_z$, $L_2=\sqrt{\lambda_2/2}
\sigma^2_z$, $L_c=\sqrt{\lambda_c/2} (\sigma^3_z+\sigma^4_z)$,
$L_r=\sqrt{\lambda_r/2} \sigma^3_z$, for positive parameters
$\lambda_\mu$, $\mu=1,2,c,r$. Here, $\lambda_1$ and $\lambda_2$ are
the strengths of the phase errors on qubits 1, 2, whereas $\lambda_c$
and $\lambda_r$ characterize the dominant (collective) and residual
(independent) dephasing on qubit 3.  For simplicity, we imagine a
situation where $|L_r|/|L_c|=\epsilon$, with $\epsilon < 1$, and
$\lambda_1 \sim \lambda_2 \sim \lambda_r \sim \lambda_0 \sim
\epsilon^2 \lambda_c$.  Physically, the collective and independent
error processes affecting qubit 3 ($a$) may or ($b$) may not have the
same origin.  In the latter case (case $b$), qubits 3 and 4 are
identically coupled to some environment, and qubit 3 is additionally
weakly interacting with a second environment. $L_c$ and $L_r$ are then
separate error generators, with an overall noise strength on qubit 3
given by $\lambda_3=\lambda_c(1+\epsilon^2)$. If the interaction
involves a single environment instead (case $a$), then the symmetry
between qubits 3 and 4 is perturbatively broken by independent errors
on qubit 3.  Accordingly, $L_c$ and $L_r$ should be combined into a
single error generator $L_3=L_c+L_r$, resulting in a noise strength
$\lambda_3= \lambda_c(1+\epsilon)^2$. In both cases, $\lambda \approx
\lambda_c$
for $\epsilon$ small enough. 

The presence of strong correlated noise naturally suggests the use of
passive QECCs as the first step toward reducing the noise. Let
$|0_L\rangle=|01\rangle_{34}$, $|1_L\rangle=|10\rangle_{34}$ define
logical DFS basis states for collective dephasing on qubits 3, 4
\cite{dfs}, and work with the new system $S'$ composed of the physical
qubits 1, 2, and the logical DFS qubit. Noise for $S'$ may be analyzed
by examining the action of the error generators on a state
$\tilde{\rho}$ that is properly initialized to $S'$. The basic
observation is that, thanks to the degenerate action of $L_c$ on DFS
states~\cite{dfss,dfs}, $L_c |i_L\rangle \langle j_L| = (\ell
\openone) |i_L\rangle \langle j_L|$ for some $\ell$ ($\ell=0$ in our
case), the errors caused by $L_c$ and $L_r$ on information encoded in
$S'$ can be described, in {\it both} cases $a$ and $b$, as
\begin{equation}
\cD_{L_c}[\tilde{\rho}] + \cD_{L_r}[\tilde{\rho}] = 
\cD_{L_r}[\tilde{\rho}] \:.
\label{encgen}
\end{equation}
Thus, the strong collective noise disappears and, since $\sigma_z^3$
acts as an encoded $\sigma_z^L$ observable, the residual noise from
$L_r$ corresponds to logical phase errors with a reduced strength. The
overall noise strength for $S'$ becomes $\lambda'=3\lambda_0$,
suitable for compensation by an additional level of QEC. In
particular, a standard three-bit QECC is able to improve the one-qubit
memory fidelity from $F_e^{\tt 1}(t) = 1-O(\lambda_0 t)$ to $F_e^{\tt
qec,1}(t) = 1- O[(\lambda_0 t)^2]$.

Some generalizations are worth mentioning. For hybrid error models
where strong correlated dephasing coexists with weak {\it arbitrary
single-qubit errors}, concatenation between an ``inner'' DFS coding on
the appropriate pairs and an ``outer'' five-bit QECC is applicable. If
the strong noise involves {\it arbitrary collective errors}, then
three-bit NSs~\cite{klv,ns} offer the most efficient code to be used
at the lowest level. In this case, the analysis is easier under the
assumption that the residual noise is in the commutant~\cite{klv} of
the primary collective generators. This assumption, which parallels
the {\it no-leakage} assumption of standard concatenated
coding~\cite{knill:1996}, ensures that the residual errors for the
logical subsystem can be, as above, described in terms of encoded
observables. Together with the identity action of the collective
generators on the NS~\cite{klv}, this implies (similar to
Eq. (\ref{encgen})) that the strong noise is fully compensated for
after the first stage of encoding.  Concatenation with active QEC can
then further suppress errors. The no-leakage assumption may be relaxed
at the expense of complicating the error control strategy.  The
consequences of leakage appear at first more serious in the
perturbative scenario $a$, as the action of $\cD_L$ would include,
besides (\ref{encgen}), additional terms mixing $L_c$ and $L_r$ or
order $\epsilon$~\cite{dfss}. However, general stability results
\cite{bacon,zanardi3} ensure that the contribution from these terms to
{\it all} the error rates $1/\tau_k^k$ remains of order $\epsilon^2$,
as they are already in the independent scenario $b$. This makes
concatenation with QEC still advantageous, provided that the procedure
is modified to detect and handle leakage
appropriately~\cite{lidar2,hwang,byrd04}.

\begin{figure}
\includegraphics[width=8cm,height=3.5cm]{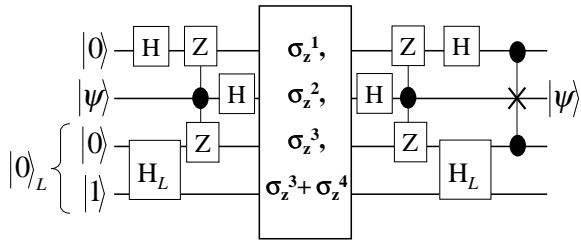}
\caption{{\bf Quantum network for the DFS-QEC code.}  Top
to bottom: Physical ancilla qubit, 1; data qubit, 2; Logical ancilla
qubit, made up by physical qubits 3 and 4.  $H_L$ denotes a logical
Hadamard operation~\cite{dfs}.  The applied error model implemented
incoherent independent $z$ noise on qubits $1$, $2$, and $3$,
superposed to a collective $z$ noise on qubits $3$ and $4$.
\vspace*{-1mm}}
\label{network}
\end{figure}

{\it Experimental implementation.-} The experiment implemented the
above-mentioned DFS-QEC code for hybrid phase errors on four qubits
using liquid-state NMR techniques~\cite{review}.  A 400 MHz Bruker
{\sc avance} spectrometer was used with a sample of $\mbox{}^{13}$C
labeled crotonic acid in a deuterated acetone
solvent~\cite{7qubitBench}. The experiment combined basic steps used
in the implementation of both active~\cite{qec,sharf} and passive
QECCs~\cite{ns,dfs}. The quantum network is shown in
Fig.~\ref{network}.
The required independent and collective errors were engineered by
combining the action of unitary radio-frequency pulses with the
non-unitary dynamics induced by magnetic field gradients integrated
over the three independent directions of the spatially distributed
sample~\cite{kspace}. If $k_i$, $i=1,2,3$, is the wavenumber of the
gradient ramp along the $i$th axis~\cite{kspace}, the phase coherence
of the corresponding qubit averaged along that direction is attenuated
by a factor $\text{sinc}(k_i(t) {L_i}/{2})$, $k_i(t)=\gamma G_i t$,
with $L_i$, $\gamma$, $G_i$, and $t$ denoting the length of the sample
in the $i$th direction, the gyromagnetic ratio of the nuclear species,
the gradient strength and the duration of the gradient pulse,
respectively.  For independent errors, the gradients were calibrated
to yield equal noise strengths on the different qubits that is,
$k_i(t) L_i =k_0(t) L$, $i=1,2,3$, for an effective length $L$. For
collective errors, additional gradients were applied on qubits 3 and 4
so that $k_3=k_c+k_0$, corresponding to a single environment situation
(case $a$) with $k_0/k_c \sim 0.5$.  In order to minimize the impact
of natural decoherence, error models corresponding to a different
noise strength were engineered by varying the value of $G_i$ while
keeping the total length of the experiment fixed.

Four different scenarios were investigated. First, the regular
three-qubit QECC with independent $z$ noise on qubits 1, 2, 3, was
implemented as a reference for the concatenated DFS-QEC code.  Second,
strong $z$ noise on qubit 3 was added.  Because, due to the incoherent
nature of the applied noise, the attenuation of the phase coherence
still yields an initial zero slope even in the absence of QEC, we also
applied the error model to different input states with no QEC.  In the
fourth scenario, the four-qubit concatenated DFS-QEC code was
realized, starting from the conditional pseudo-pure
states~\cite{review,q5} $ \mket{0}^{1}\mbra{0} \otimes \sigma_u^{2}
\otimes \mket{0}^{3}\mbra{0}\otimes \mket{0}^{4}\mbra{0}$, $u=x$, $y$,
$z$.  Ideally, the data qubit should be recovered with the same
accuracy as in the original QEC setting.  Four- and one-qubit state
tomography~\cite{Chuang-Tomo} was performed to verify both input and
output states, as well as their correlations. A total of $18$ readout
pulses was used to reconstruct the four-qubit state while two pulses
sufficed for the state of the data qubit $2$ alone.  Strongly
modulating control pulses designed to be robust against
radio-frequency power inhomogeneity were employed~\cite{EvanPulses}. A
feedback loop was implemented to correct for systematic errors arising
from the response of the electronics chain~\cite{BoulantEntSwap}.  The
protons were decoupled during both the experiment and the acquisition
to avoid additional incoherence.
 
{\it Results.-} For an incoherent error dynamics as implemented in 
the experiment, the behavior of the error-corrected entanglement 
fidelity for equally distributed, independent phase errors with 
strength $k_0$ is
\begin{eqnarray*}
F_e^{\tt qec,1}(t)= {1 \over 2} + {1 \over 4} 
\Big( 3\, \text{sinc}(k_{0}(t){L}/{2}) -
\text{sinc}^3(k_0(t){L}/{2}) \Big) \,.
\end{eqnarray*} 
When the strong noise component is added to qubit 3, the above 
equation is modified as
\begin{eqnarray*}
F_e^{\tt qec,1}(t) = {1 \over 2} &\hspace{-1mm}+\hspace{-1mm} & {1
\over 4} \Big( 2\,
\text{sinc}(k_{0}(t){L}/{2})+\text{sinc}(k_3(t){L}/{2})\\
&\hspace{-1mm} - \hspace{-1mm}&
\text{sinc}^2(k_{0}(t){L}/{2})\,\text{sinc}(k_3(t){L}/{2}) \Big)\:.
\end{eqnarray*}
For irreversible Markovian noise as in Eq. (\ref{sme}), the standard
expressions for $F_e^{\tt qec,1}(t)$~\cite{sharf} are recovered upon
replacing each sinc function by a corresponding exponential.  From the
above expressions, it is clear that QEC compensates for errors to first
order, irrespective of the collective noise component $k_c$.  However,
for longer times the latter decreases $F_e^{\tt qec,1}(t)$ much faster
than the weaker independent noise, reducing the effectiveness
of the code.

\begin{figure}
\includegraphics[height=8.3cm]{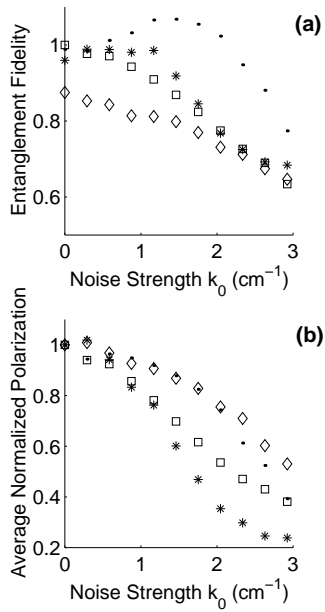}
\caption{{\bf Experimental results.}  {\bf (a)} Entanglement fidelity,
{\bf (b)} Average polarization of the output states as a function of
the applied noise strength. Dots: 3-qubit QEC with independent $z$
noise only; Asterisks: Same 3-qubit QEC under independent plus
collective $z$ noise; Squares: Pure independent $z$ noise without
QEC; Diamonds: DFS-QEC data. The error bar is $\pm0.02$ for each 
data point (not diplayed for clarity). See text for an explanation 
of the different effects. \vspace*{-2mm}}
\label{EntFidRes2}
\end{figure}

For each set of experiments, entanglement fidelities were inferred by
calculating $F_e= (C_x+C_y+C_z+1)/4$, under the assumption of unital
dynamics and with $C_u={\trace(\rho_{\tt in, u}\rho_{\tt out,
u})}/{\trace(\rho_{\tt in, u}^2)}$, $ \rho_{\tt in, u}$, $\rho_{\tt
out, u}$, $u=x,y,z$, being the measured input and output states,
respectively.  The results are summarized in Fig.~\ref{EntFidRes2}(a).
The four-qubit input state correlations with the intended states were
estimated to be on average $0.93\pm0.02$ after state tomography.  As a
first remark, QEC with independent noise alone (dots) shows some
improvement compared with the uncorrected scenario (squares).  The
noticeable "hump" in the data may be understood as a consequence of
imperfect initialization of the ancillae, combined with the
pseudo-pure state nature of the underlying NMR
states~\cite{review,humpMaterial}.  A similar effect (although less
pronounced) is present in the QEC data under additional strong
collective noise (asterisks). In any case, the efficiency of QEC in
the presence of both independent and collective errors is greatly
reduced as expected. The diamonds correspond to the DFS-QEC
concatenated code. The initial fidelity drop is primarily explained
by coherent errors associated with the longer pulse sequence necessary 
to realize the code, additional natural decoherence being induced 
as soon as qubit 4 is brought into the $xy$ plane. 
Most of these features were accounted for by extensive simulations
including both coherent and incoherent errors, as well as imperfect
readouts. In particular, values of $F_e$ larger than one arise from
the joint influence of nuclear Overhauser enhancement, coherent and
incoherent errors on the four-qubit system, along with the hump
effect mentioned above.

While the entanglement fidelity data of Fig.~2(a) provide a complete
representation of the overall implementation accuracy, the significant
impact of coherent and initialization errors evidenced by the
simulations makes it difficult to directly assess the performance of
different schemes at correcting the intended error model.  This
suggests to also analyse the data by using a metric which is only
sensitive to the length of the output states.  A natural choice is
provided by the average output polarization, $P=(P_x+P_y+P_z)/3$,
where $P_u={\trace(\rho_{\tt out, u}^2)}/ {\trace(\rho_{\tt out, u
0}^2)}$, $\rho_{\tt out, u}$, $\rho_{\tt out, u 0}$ being the output
corresponding to input $u$ with and without the applied noise,
respectively (the same metric was recently used
in~\cite{LidarNMR}). The results are shown in
Fig.~\ref{EntFidRes2}(b).  These now clearly demonstrate the
advantages of using the DFS-QEC code, if coherent errors and initial
polarization losses can be made negligible. As expected, the original
QEC behavior with independent noise only (dots) is recovered with good
accuracy by the DFS-QEC code under both independent and collective
noise (diamonds). Further analysis of the data revealed that the hump
visible in Fig.~\ref{EntFidRes2}(a) is still present but much less
important because of the insensitivity of the new metric to the
initial coherent errors in $\rho_{\tt out, u 0}$.

{\it Conclusions.-} Our work provides the first experimental instance
of a concatenated QEC code, using a combination of both physical and
lower-level logical qubits to stabilize the quantum data at a
higher logical level. While our results point to the need for
improved control capabilities for both unitary and non-unitary
dynamics, the implementation convincingly shows the benefits of
concatenated active and passive QECCs in handling realistic error
models. Notably, collective phase errors are the limiting factor for
reliable storage using trapped ions~\cite{kielpinski}. Thus, our
results might allow further advances toward realizing fault-tolerance
in scalable device technologies.

{\it Acknowledgments.-} We thank Rudy Martinez for synthesizing the
crotonic acid sample. Work supported by ARDA and ARO.  L.V.
acknowledges support from MIT and LANL during the early stages of 
this project.

\end{document}